
\input amstex
\documentstyle{amsppt}
\NoBlackBoxes
\def\ss{\scriptscriptstyle}
\topmatter
\date August 1, 1995\enddate
\title
Spinors, Jets, and the Einstein Equations
\endtitle
\author C. G. Torre\endauthor
\affil Dept.~of Physics, Utah State University, Logan, UT 84322-4415,
USA\endaffil
\email torre\@cc.usu.edu\endemail
\leftheadtext{C. G. Torre}
\rightheadtext{Spinors, Jets, and the Einstein Equations}
\abstract
Many important features of a field theory, {\it e.g.}, conserved
currents,
symplectic structures, energy-momentum tensors, {\it etc.}, arise as
tensors
locally constructed from the fields and their derivatives.  Such tensors
are
naturally defined as geometric objects on the jet space of solutions to
the
field equations.   Modern results from the calculus on jet bundles can
be
combined with a powerful spinor parametrization of the jet space of
Einstein metrics to unravel basic features of the Einstein equations.
These
techniques have been applied to computation of generalized symmetries
and
``characteristic cohomology'' of the Einstein equations, and lead to
results
such as a proof of non-existence of ``local observables'' for vacuum
spacetimes and a uniqueness theorem for the gravitational symplectic
structure.
\endabstract
\endtopmatter

\document

\head 1\enspace Introduction\endhead

This is a survey of results of work performed largely in collaboration
with
Ian Anderson (Dept.~of Mathematics, Utah State University).  The
presentation will
be somewhat informal; rigorous statements and proofs of our results will
be presented elsewhere \cite{Torre and Anderson (1993)}, \cite{Anderson
and Torre (1994)}, \cite{Anderson and Torre (1995)}.  In the very
broadest
terms, our efforts are intended to help answer the question:  ``In what
ways
are the Einstein equations special?''  There are of course a plethora of
special features of the Einstein equations which have been uncovered
since
the advent of general relativity.  Here are some examples.  The
vacuum equations in 4 dimensions are, up to specification of the
cosmological constant, the only second order partial differential
equations
one can write down for a metric which are ``generally covariant'' and
can
be derived from a variational principle.  Despite the complexity of the
field
equations, a large number of exact solutions are known.  Indeed, certain
reductions of the equations (self-dual equations, stationary-
axisymmetric
vacuum and electrovac equations) admit transitive symmetry groups and
are in some sense ``integrable''.  Special features that are of
particular
relevance to the physical viability of the Einstein equations include
theorems guaranteeing that the Cauchy problem is well posed, existence
and uniqueness (up to diffeomorphisms) of solutions, existence and
positivity of conserved energy-momentum in the asymptotically flat
context, {\it etc}.  Finally, a feature of the Einstein equations which
is
especially
relevant for attempts at quantization is that the equations constitute a
constrained Hamiltonian system.

The modern geometric theory of differential equations provides
systematic means through which one can characterize certain important
structural
features of any set of field equations.  These structural features arise
as
geometric objects on the {\it jet space} of solutions to the field
equations \cite{Saunders (1989)}, \cite{Olver (1993)} and are neatly
analyzed in terms of the {\it variational
bicomplex} \cite{Anderson (1992)}, \cite{Olver (1993)}.  Such jet space
techniques have been used in the analysis
of a number of differential equations of applied mathematics and
mathematical physics.  It is now
possible to apply these techniques to study special features of the
vacuum Einstein equations; the results of such investigations are the
subject of this paper.  In particular, I will report on a systematic
classification of {\it generalized symmetries} and {\it
generalized conservation laws} for the vacuum equations.  Roughly
speaking, generalized symmetries are infinitesimal transformations
constructed locally from the metric and its derivatives mapping
solutions
to
solutions.  Generalized conservation laws are closed but not exact
differential forms constructed locally from solutions to the field
equations
and/or
solutions to the linearized equations.  The generalized conservation
laws
correspond to conserved volume, surface, or line integrals associated to
solutions to the Einstein equations.  In the mathematical literature,
the
generalized conservation laws go by the name ``characteristic
cohomology'' of the field equations \cite{Bryant and Griffiths (1993)}
(for a BRST approach to characteristic cohomology see \cite{Barnich {\it
et
al} (1994)}).

To begin our survey of generalized symmetries and conservation laws for
the Einstein equations, we should first improve our understanding of the
jet
space of solutions to the vacuum equations.

\head 2\enspace Jet Space\endhead

The jet space of solutions to the Einstein equations can be viewed as
the
beginning of an answer to the question ``What data are freely
specifiable
{\it at a point} of a vacuum spacetime?''  This question is similar in
spirit
to the kind of question one
asks when investigating the Cauchy problem, where one seeks the data
which can be freely specified on a Cauchy surface.  Of course the two
questions are quite different mathematically but, to carry the analogy a
little further, one can think of the jet space of solutions as something
of
an
analog of Cauchy data for analytic solutions.  Indeed, to construct an
analytic solution to the Einstein equations, one can pick coordinates
$x^i$
about some point $x^i=0$ and write down a power series expansion:
$$
g_{ij}(x) = g_{ij}(0) + g_{ij,k}(0) x^k + {1\over2!}g_{ij,kl}(0)x^kx^l +
\cdots.\tag 1
$$
Evidently, formal power series expansions of a metric about a point
$x^i$
are determined by the data
$$
(x^i, g_{ij}, g_{ij,k}, g_{ij,kl}, \ldots).\tag 2
$$
Such data define a point in the {\it jet space of metrics} $\Cal J$.  Of
course, to define a metric via power series about some point, one should
enforce some convergence criterion on the metric and its derivatives at
the
given point; the jet space is, however, defined without such convergence
criteria.  A more precise way to introduce the jet space of metrics is
to
view it as a bundle whose base space is spacetime and whose typical
fiber
is
the space of values of the metric and its derivatives at a point.  A
metric
defines a cross section of this bundle.  Moreover,
Borel's theorem (see, {\it e.g.}, \cite{Kahn (1980)}) implies that given
a
point $
(x^i, g_{ij}, g_{ij,k}, g_{ij,kl}, \ldots)\in
\Cal J$, there is always a smooth metric which has that data.

Let us remark that while we have defined $\Cal J$ using coordinates and
coordinate derivatives of the metric, the jet space is in fact a
coordinate-independent object.  It is possible to give a coordinate-free
 definition of
$\Cal J$ as a fiber bundle over spacetime along the lines mentioned
above.
In
particular, it is possible to replace coordinates and coordinate
derivatives
with globally defined derivative operators (see, {\it e.g.}, \cite{Wald
(1990)}), but for simplicity we shall
stick with an
informal local treatment.

Having defined the jet space of metrics we would now like to see what
points in jet space are allowed by the Einstein equations.  To do this,
we
need a slightly better parametrization of $\Cal J$.  To this end, given
coordinates $x^i$, define variables $\Gamma^{(k)}$ and $Q^{(k)}$,
where
$$
\Gamma^{(k)}\longrightarrow\Gamma^i_{j_0j_1j_2\cdots j_k} :=
\Gamma^i_{(j_0j_1,j_2\cdots j_k)},\qquad k=1,2,\ldots\tag 3
$$
and
$$
Q^{(k)}\longrightarrow Q_{ab,c_1c_2\cdots
c_k}:=g_{am}g_{bn}\nabla_{(c_3}\cdots \nabla_{c_k}
R^{m\phantom{c_1}n}_{\phantom{m}c_1\phantom{n}c_2)}\qquad
k=2,3,\ldots,\tag 4
$$
where $\nabla$ is the torsion-free derivative operator compatible with
$g_{ab}$ and $R_{abcd}$ is the Riemann tensor.  It can be shown that
$\Gamma^{(k)}$ and $Q^{(k)}$
are algebraically independent at any given point of spacetime.  The
variables $\Gamma^{(k)}$ carry the coordinate-dependent information in
the $k^{th}$ partial derivatives of the metric.  In particular, all of
the
variables
$\Gamma^{(k)}$ vanish at the origin of a geodesic coordinate chart.
The variables $Q^{(k)}$ contain all
spacetime-geometric information in the $k^{th}$ partial derivatives of
the
metric.  In particular, the curvature tensor and all of its covariant
derivatives can be uniquely expressed in terms of the variables
$Q^{(k)}$.
The tensors $Q^{(k)}$ were apparently introduced by Penrose
\cite{Penrose
(1960)} and are closely related to
Thomas's ``normal metric tensors'' \cite{Thomas (1934)}.

Our first result is that the variables
$$
(x^i, g_{ij}, \Gamma^{(1)},\Gamma^{(2)},\ldots,Q^{(2)},Q^{(3)},\ldots)
\tag 5
$$
uniquely parameterize points in the jet bundle.  In other words, given
the
data (5) one can reconstruct the metric and all of its derivatives at
the
point
labeled $x^i$.  The data (5) are freely specifiable at a point of a
(pseudo-) Riemannian manifold.

The Einstein tensor can be viewed as a collection of functions on $\Cal
J$,
and the Einstein equations can be viewed as defining a subspace
(actually a
submanifold) ${\Cal E}\hookrightarrow {\Cal J}$.  Because the Einstein
equations are geometrically defined, they introduce
relations only among the variables $Q^{(k)}$.  In fact, the vacuum
equations uniquely fix
the traces of these tensor in terms of their trace-free parts
\cite{Anderson and Torre (1994)}.  Thus
a point in the jet space of Einstein metrics is defined by the variables
$$
(x^i,g_{ij},\Gamma^{(1)},\Gamma^{(2)},\ldots,\widetilde
Q^{(2)},\widetilde Q^{(3)},\ldots),\tag 6
$$
where $\widetilde Q^{(k)}$ denotes the completely trace-free part of
tensors (4) with respect to the metric $g_{ij}$.  The data (6) are
freely
specifiable at a point of a Ricci-flat spacetime.  These coordinates on
$\Cal E$ can be interpreted in terms of a power series expansion
of an Einstein metric as follows.  If we are trying to build an
Einstein
metric by Taylor series we (i) specify the spacetime point $x^i$ around
which the series is being developed, (ii) specify the metric
components $g_{ij}$ at $x^i$, (iii) specify the variables
$\Gamma^{(k)}$;
this
fixes the coordinate system in which the metric is being built, (iv)
specify
the variables $\widetilde Q^{(k)}$, which supplies the geometric content
of
the Einstein metric.  Of course, this procedure leaves open the question
of
convergence of the series.

The parametrization (6) of $\Cal E$ turns out to be somewhat unwieldy in
applications,
primarily because of the need to remove so many traces.  A much more
useful parametrization, which only exists in four dimensions, uses a
spinor
representation of the variables
$\widetilde Q^{(k)}$.  Let $\Psi_{\ss ABCD}$ and $\overline\Psi_{\ss
A^\prime B^\prime C^\prime D^\prime}$ denote the Weyl spinors
\cite{Penrose (1960)}.  Fix a
soldering form $\sigma_a^{\ss AA^\prime}$ such that, for a given
$g_{ij}$,
$$
g_{ij}=\sigma_i^{\ss AA^\prime}\sigma_{j\ss AA^\prime}.\tag 7
$$
It can be shown that the variables $\widetilde Q^{(k)}$ are uniquely
parametrized by
the soldering form, the spinor variables
$$
\Psi^{(k)}\longleftrightarrow \Psi{}_{\ss J_1\cdots J_{k+2}}^{\ss
J_1^\prime \cdots J_{k-2}^\prime}
=\nabla_{\ss (J_1}^{\ss( J_1^\prime}\cdots\nabla_{\ss J_{k-2}}^{\ss
J^\prime_{k-2})}\Psi_{\ss J_{k-1}J_kJ_{k+1}J_{k+2})},\tag 8
$$
and their complex conjugates $\overline\Psi^{(k)}$.
Thus we obtain a spinor parametrization of $\Cal E$
in terms of
$$
(x^i,g_{ij},\Gamma^{(1)},\Gamma^{(2)},\ldots,
\Psi^{(2)},\overline\Psi{}^{(2)},\Psi^{(3)},\overline\Psi{}^{(3)},\ldots
).
\tag 9
$$
The spinor aficionado will recognize that our spinor parametrization of
$\Cal E$ is closely related to Penrose's notion of an ``exact set of
fields'' \cite{Penrose (1960)}.

The parametrization (9) of the jet space of solutions to the Einstein
equations is an important technical tool needed to classify
symmetries and conservation laws.  More generally, these
variables allow us to address problems of the following type.  Suppose
we
are
interested in finding a tensor field $T=T(x,g,\partial g,\ldots)$,
locally
constructed from the metric and its derivatives to some order, which
satisfies some local differential relations,
$$
DT=0,\tag 10
$$
when the Einstein equations hold.  As an example, suppose we wanted to
find a conserved current for the Einstein equations.  This would be a
vector field $j^a=j^a(x,g,\partial g,\ldots)$ such that
$$
\nabla_aj^a = 0\qquad \text{when $G_{ab}=0$}.\tag 11
$$
The tensors $T$ and $DT$ can be viewed as a collection of functions on
$\Cal J$ and the differential relation (10) says that the function $DT$
vanishes
on $\Cal E$.   If we express the relation (10) in the coordinates (9),
then
(10)
must hold {\it identically}.  The power of spinor analysis can now be
brought to bear on classifying all such solutions $T$ to this identity
up
to
terms which vanish on $\Cal E$.  As we
shall see, classifications of generalized symmetries and generalized
conservation laws are precisely problems of this type.

To be honest, it is a
bit of an over-simplification to say that these problems can be solved
in a
straightforward manner given the coordinates (9).  The results to be
given
below are in fact intricately related through a mathematical structure
on
$\Cal J$ (or $\Cal E$) called ``the variational bicomplex''
\cite{Anderson
(1992)}.  Because my
goal here is to emphasize results rather than techniques, I will not be
able
to say more about the variational bicomplex in this paper.  Suffice it
to
say
that the bicomplex is an indispensable tool in the analysis of any set
of
field
equations, and this mathematical structure is playing a vital role
``behind
the scenes'' in all that follows.

\head 3\enspace Generalized Symmetries\endhead

A generalized symmetry is an infinitesimal transformation constructed
locally from the relevant fields and their derivatives to some order
which
maps solutions of the field equations to other solutions.  Generalized
symmetries are of interest because they provide methods of generating
new
solutions from known solutions, their existence is necessary for the
existence of local conservation laws, and because of their role in
complete
integrability of a variety of partial differential equations \cite{Olver
(1993)}.  Before giving results from our
classification of generalized symmetries of the Einstein equations, let
us
first have a look at an elementary example of a dynamical system that
admits a generalized symmetry.

Consider the Kepler problem, which can be described by the non-linear
system of ordinary differential equations:
$$
{\bold r}^{\prime\prime}=-k{{\bold r}\over r^3},\tag 12
$$
where ${\bold r}$ is the relative position of two masses in space, $k$
is a
constant, and a prime denotes a time derivative.  A point in the jet
space
$\Cal J$ for this problem is defined by the variables $(t, {\bold r},
{\bold
r}^\prime,{\bold r}^{\prime\prime},\ldots)$.  The equations (12) and
their
time derivatives define the
jet space ${\Cal E}\hookrightarrow{\Cal J}$ of solutions.  Coordinates
on
$\Cal E$ are $(t,{\bold r}, {\bold r}^\prime)$; these variables
parametrize
the extended velocity phase space for the Kepler problem.
There are a
couple of obvious symmetries of the equations (12), namely, rotational
symmetry and time translation symmetry.  These symmetries are usually
called ``point symmetries'' or ``Lie symmetries''.  The point symmetries
are distinguished by the fact that they can be defined as groups of
transformations of the underlying space of independent and dependent
variables $(t, {\bold r})$ only, without involving derivatives of $\bold
r$.
By contrast, the following infinitesimal transformation necessarily
involves
derivatives of ${\bold r}$:
$$
\delta{\bold r}=2({\bold \lambda}\cdot{\bold r}){\bold r}^\prime
-({\bold \lambda}\cdot{\bold r}^\prime){\bold r}-({\bold r}\cdot{\bold
r}^\prime){\bold \lambda}.\tag 13
$$
Here $\bold \lambda$ is a fixed, time-independent vector.  It is
straightforward to check that if ${\bold r}(t)$ satisfies (12) then, to
first
order
in $\bold \lambda$, ${\bold r}(t) + \delta{\bold r}(t)$ also satisfies
(12).  The
transformation (13) represents a first-order generalized symmetry of the
equations (12)\footnote{Note that all generalized symmetries of (12) can
be
expressed as
first-order symmetries if we use the equations of motion.  This
property of generalized symmetries does not generalize to partial
differential equations}.
I think you
will
agree that it is somewhat remarkable that the relatively simple system
of
equations (12) admits such a complicated ``hidden symmetry''.  The
three-parameter family of symmetries given in (13) are responsible for
the
existence of a  conserved vector, known as the Laplace-Runge-Lenz vector
${\bold A}$ (see, for example, \cite{Goldstein(1990)}):
$$
{\bold A} = {\bold r}^\prime\times({\bold r}\times{\bold r}^\prime)-
k{{\bold
r}\over r}.\tag 14
$$
Conservation of the Laplace-Runge-Lenz vector reflects a special feature
of
the
Kepler problem:  all its bound orbits are closed.  The only other
central
force that has this special property is that of an isotropic oscillator,
and in
this case there is again a generalized symmetry and (tensor)
conservation
law which is responsible.

It is natural to wonder if the Einstein equations will admit any hidden
generalized
symmetries as do many simpler systems of non-linear differential
equations.  To find such symmetries we must look for an infinitesimal
transformation
$$
\delta g_{ab} = h_{ab}(x, g, \partial g,\ldots)\tag 15
$$
mapping solutions to solutions.  This means that $h_{ab}$ must satisfy
the
linearized equations,
$$
-\nabla^c\nabla_c h_{ab} - \nabla_a\nabla_b (g^{cd}h_{cd})
+2\nabla^c\nabla_{(a}h_{b)c}=0,\tag 16
$$
when the metric $g_{ab}$, out of which $h_{ab}$ is built, satisfies the
Einstein equations $G_{ab}=0$.  Because we are only interested in
transformations between solutions, any two symmetries that are equal
when
the field equations hold can be considered equivalent.  So, for example,
the Einstein tensor defines a generalized symmetry, $h_{ab}=G_{ab}$
but the
transformation it generates is trivial, {\it i.e.}, we identify this
symmetry with $h_{ab}=0$.  As described earlier,
the classification of on-shell generalized symmetries is exactly the
kind
of problem that can be fruitfully attacked
using the spinor parametrization of $\Cal E$.  In detail, the symmetry
transformation $h_{ab}$ is viewed as a function on $\Cal J$ (we are
actually only interested in the restriction of $h_{ab}$ to $\Cal E$).
We
view
(16) as requiring that
certain functions on $\Cal J$ built from $h_{ab}$ vanish when restricted
to
$\Cal E$.  That is, as
a function on $\Cal E$, $h_{ab}$ should satisfy the identity (16).  As
an
identity on $\Cal E$, (16) must hold for all values of the
coordinates (9); analysis of this requirement leads to the following
result.

Let $h_{ab}=h_{ab}(x, g, \partial g,\ldots)$ be a generalized symmetry
of
the vacuum Einstein equations in four spacetime dimensions.  Then there
is
a constant $c$ and a covector $V_a=V_a(x,g,\partial g,\ldots)$ such
that,
modulo terms that vanish when $G_{ab}=0$, the symmetry must take the
following form:
$$
h_{ab}= c\,g_{ab} + \nabla_a V_b + \nabla_b V_a.\tag 17
$$
This form of $h_{ab}$ represents a combination of two types of symmetry
transformations.  The term $c\,g_{ab}$ corresponds to a scale symmetry
admitted by the vacuum equations.  If we allowed for a cosmological
constant, this symmetry would be absent.  The terms
$\nabla_a V_b + \nabla_b V_a$ correspond to the infinitesimal change in
the metric arising from the pull-back by a 1-parameter family of (local)
spacetime diffeomorphisms generated by $V^a$.  Thus the most general
symmetry is a combination of a constant scaling and a ``gauge
transformation''.  Of course, both of these symmetries are well-known
and
we conclude that the vacuum Einstein equations admit no ``hidden local
symmetries''.

\head 4\enspace Generalized Conservation Laws\endhead

The symmetries we have found do not have any non-trivial conservation
laws associated with them.  Noether's theorem then suggests that,
aside from
possible topological conservation laws, there are
no conserved currents for the Einstein equations built locally from the
metric and its derivatives.  This is in fact true, but it does not
follow
directly from our symmetry classification because it is {\it a priori}
possible to have symmetries that are on-shell trivial and nevertheless
generate non-trivial conservation laws.  For completely
non-degenerate systems of PDE's it is known that every non-trivial
conserved current
follows from a non-trivial generalized symmetry \cite{Olver (1993)}, but
the Einstein
equations do not qualify as a completely non-degenerate system owing to
their general covariance.  The problem of rigorously classifying
conserved
currents for the vacuum Einstein equations can be solved using the
variational
bicomplex and our spinor techniques.  In fact, it is possible to
generalize
the analysis and classify all closed forms locally built from vacuum
metrics
as well as a large class of closed forms locally built from vacuum
metrics
and solutions of the linearized equations.  We begin with closed forms
built
locally from vacuum metrics (see \cite{Wald (1990)} for a general
discussion of identically closed forms locally built from fields).

Let $\omega=\omega(x,g,\partial g,\ldots)$ be a $p$-form locally built
from the metric and its derivatives to some order.  We assume $p<4$.  If
$d\omega=0$ when the vacuum Einstein equations hold, we say that
$\omega$ represents a {\it generalized conservation law} for the vacuum
equations.  The importance of a generalized conservation law stems from
the
fact that
its integral over a closed\footnote{With suitable boundary conditions,
the
submanifold can be open or have boundaries.} $p$-dimensional
submanifold $\Sigma$ is independent of the choice of $\Sigma$ (up to
homology) when the metric satisfies the field equations.  Thus the
integral
$$
Q[g] = \int_{\Sigma}\omega(x,g,\partial g,\ldots)\tag 18
$$
is a conserved charge characterizing vacuum spacetimes.  As we are only
interested in the values of the conserved charges for solutions of the
field
equations, we will identify any generalized conservation laws which are
equal when the field equations hold.  In other words, viewing the
generalized conservation laws as functions on $\Cal J$, we are only
interested in their restriction to $\Cal E$.

Of course, if (on $\Cal E$) there exists a $(p-1)$-form
$\eta=\eta(x,g,\partial g,\ldots)$
such that
$$
\omega=d\eta,\tag 19
$$
then $\omega$ is identically closed and $Q[g]=0$ for any metric $g$.  We
will call such exact forms {\it trivial conservation laws}.

If $\omega$ is a closed 3-form, then its Hodge dual is a conserved
current;
the
conserved charge is then a volume integral of a local density.  In
field theory this is the way in which conserved charges typically arise.
However, interesting
conserved quantities do sometimes arise not as volume integrals but
instead
as
surface---or even line---integrals.  For example, if we restrict our
attention
to vacuum (regions of) spacetimes admitting a Killing vector field
$k^a$,
then the Komar 2-form,
$$
\kappa_{ab}=\epsilon_{ab}{}^{cd}\nabla_ck_d,\tag 20
$$
defines a generalized conservation law, as does the twist 1-form
$$
\tau_a=k^b\kappa_{ab}.\tag 21
$$
Is there a field-theoretic explanation for the existence of these
generalized
conservation laws?  Are there any generalized conservation laws admitted
by the vacuum Einstein equations which can be defined without assuming
the existence of a Killing vector?  We shall answer the latter question
now
and return to the first question toward the end of this article.

We wish to classify closed forms locally built from Ricci-flat metrics.
Once again we are looking for some functions on $\Cal J$ satisfying a
certain differential relation on $\Cal E$.  Using spinor techniques and
basic
properties of the variational bicomplex we obtain the following results.
If
$\omega=\omega(x,g,\partial g,\ldots)$ is a $p$-form locally constructed
from the metric, and $\omega$
is closed when the Einstein equations hold then, modulo terms which
vanish
when the field equations are satisfied, there exists a $(p-1)$-form
$\eta=\eta(x,g,\partial g,\ldots)$ and a constant $c$ such that
$$
\eqalign{
\omega &= \text{const.},\qquad\ \ \  p=0;\cr
\omega&=d\eta,\qquad\quad\ \ \ \ \, p=1,2;\cr
\omega&=c\,\sigma + d\eta,\qquad p=3.}\tag 22
$$
In (22) $\sigma$ is an {\it identically} closed $3$-form locally built
from
the
metric and its first derivatives.  This $3$-form is a representative of
the
non-trivial cohomology class at degree 3 which exists on the bundle of
Lorentzian metrics over spacetime \cite{Torre (1995)}.  The
corresponding
conserved
charge is the ``kink-number'' of the spacetime, which was discussed by
Finkelstein and Misner quite some time ago \cite{Finkelstein and Misner
(1959)}.  Roughly speaking, the
kink
number counts the number of times light cones tumble as one traverses a
hypersurface.  Because $\sigma$ is identically closed, this conservation
law
exists for any gravitational theory in which one employs a Lorentzian
metric.  Aside from this single topological conservation law, the vacuum
Einstein equations admit no non-trivial generalized conservation laws.

There is an immediate corollary of this result which is relevant for the
Hamiltonian formulation of general relativity on closed universes
\cite{Torre(1993)}.
Recall that the Einstein
equations can be viewed as defining a constrained Hamiltonian system.
The
constraints limit the possible range of canonical data and generate
canonical transformations which are Hamiltonian expressions of the
action
of spacetime diffeomorphisms on the Cauchy data.  It is of interest,
especially in attempts to canonically quantize the theory, to find
functions
on the gravitational phase space which are invariant under these
canonical
transformations, {\it i.e.}, which have vanishing Poisson brackets with
the
constraint functions.  These ``gauge invariant'' functions on phase
space
are commonly
called the {\it observables} of
the theory.  Because, for closed universes, the Hamiltonian of general
relativity is a linear
combination of the constraint functions, the
observables are constants of the motion.  But we have, in effect, just
classified a large set of constants of motion for the Einstein
equations,
and
so we have a classification of a particular type of observables.

Let us define a {\it local observable} $\Cal O$ as an observable on the
gravitational phase space which is constructed as an integral over a
compact
$p$-dimensional
manifold $\Sigma$ of a differential $p$-form $\omega$,
where $\omega$ is
locally built from the canonical coordinates and momenta and their
derivatives to any finite order.  Because the canonical
variables are related by local formulas to the spacetime metric it
follows
that, on solutions (in which $\Sigma$ is embedded in spacetime), $\Cal
O$
is
a conserved charge of the type (18).  But there are no such conserved
charges
except the kink number, which must always vanish in spacetimes that
admit
a
foliation by spacelike hypersurfaces \cite{Finkelstein and Misner
(1959)}.  We conclude that there are no
local
observables for vacuum spacetimes with compact Cauchy surfaces.  This is
not to say that
there are no observables, but only that the simplest class of
observables
that
one naturally tries to construct for general relativity are simply not
there.

Let us now turn to generalized conservation laws built from solutions
$\gamma_{ab}$ to the linearized Einstein equations.  An important
example is provided by the pre-symplectic current for the vacuum
equations.  Here we present it as a
3-form:
$$
\Omega_{abc}=\left(\gamma^{\,m}_{[a}\nabla^n\hat\gamma^{ p}_b
-\hat\gamma^{ m}_{[a}\nabla^n\gamma^{\,p}_b\right)
\epsilon_{c]mnp}.\tag 23
$$
The pre-symplectic current is locally built from the metric, a pair of
metric perturbations $\gamma$ and $\hat\gamma$, and their first
derivatives.  $\Omega$ is closed when the metric and
perturbations satisfy the Einstein equations and linearized equations
respectively.  Thus its integral over a compact
hypersurface,
$$
\Xi(\gamma,\hat\gamma)=
\int_\Sigma \Omega,\tag 24
$$
is independent of the choice of $\Sigma$ up to homology${}^2$.  Up to a
normalization, $\Xi(\gamma,\hat\gamma)$ is the value of the
pre-symplectic
form on the space of solutions to the vacuum Einstein equations
when acting on a pair of tangent vectors $(\gamma,\hat\gamma)$ (see, for
example, \cite{Ashtekar (1990)}, and references therein).

Integrable systems of PDE's often admit inequivalent symplectic
structures
\cite{Olver (1993)}.  In addition, it is known that the existence of
generalized conservation
laws depending on three or more solutions to the linearized equations is
closely related to applicability of Darboux's method of integration
\cite{Anderson (1992)}.  Thus
this sort of conservation law---depending on one or more solutions to
the
linearized equations--- can expose important features of a set of field
equations, and it is natural to ask if the Einstein equations admit any
other
generalized conservation laws of this type.

To present our classification of generalized conservation laws depending
on
solutions to the linearized equations we must introduce some notation.
Let
$\omega^{(p,q)}$ denote a spacetime $p$-form locally constructed from
the metric and its derivatives as well as from $q$ solutions of the
linearized
equations and their derivatives.  The linearized solutions must appear
in a
skew $q$-multilinear fashion.  For example, the pre-symplectic current
$\Omega_{abc}$ would be denoted $\omega^{(3,2)}$ in this notation, and
the forms classified in (22) would be denoted $\omega^{(p,0)}$.  We
now look for such forms that are closed when the Einstein equations and
their linearization are satisfied by the metric and perturbations.  To
find
such forms, we must generalize our spinor parametrization of $\Cal E$ to
include the jet space of solutions to the linearized equations, but this
is
relatively straightforward.  Using techniques from the variational
bicomplex we obtain the following results.

Let $\omega^{(p,q)}$, $q>0$ and $p<4$, be a generalized conservation
law
for the
vacuum Einstein equations in four spacetime dimensions.  Then there
exist
constants $b$ and $c$ and forms $\eta^{(p-1,q)}$ locally constructed
from
the
metric and perturbations such that,
modulo terms which vanish when the Einstein and linearized Einstein
equations hold, we have:
$$
\eqalign{
\omega^{(0,q)}&=0;\cr
\omega^{(p,q)}&=d\eta^{(p-1,q)},\qquad p=1,2;\cr
\omega^{(3,q)}&=d\eta^{(2,q)},\qquad\quad q>2;\cr
\omega^{(3,2)}&=b\,\Omega + d\eta^{(2,2)};\cr
\omega^{(3,1)}&=c\,\Theta + d\eta^{(2,1)}.}\tag 25
$$
In this last equation we have denoted by $\Theta$ a 3-form
$\Theta_{abc}$, which is defined as
$$
\Theta_{abc}=\epsilon_{abc}{}^d\nabla^e\left(\gamma_{de}-
g_{de}g^{mn}\gamma_{mn}\right).\tag 26
$$
$\Theta$ is closed when $g_{ab}$ and $\gamma_{ab}$ satisfy the Einstein
equations and their linearization respectively.
The integral of $\Theta$ over a hypersurface is again independent of the
choice of hypersurface and defines the canonical 1-form, or
pre-symplectic
potential, on the space of solutions to the Einstein equations.  This
means
that, viewing the conserved charge defined by $\Theta$ as a 1-form on
the
infinite-dimensional space of solutions acting on a tangent vector
$\gamma$, the exterior derivative of the
charge is the symplectic 2-form on the space of solutions.  Aside from
the
symplectic current and its associated potential, there are no other
non-trivial generalized conservation laws built from solutions of the
linearized equations as described above.  Note that this result
establishes
a
uniqueness theorem for the gravitational pre-symplectic structure in the
sense that any such structure which can be constructed as the spatial
integral of a closed, locally constructed 3-form is a multiple of
(24).

\head 5\enspace Discussion\endhead

We have classified the generalized symmetries and generalized
conservation laws of the vacuum Einstein equations in four dimensions.
Our results indicate that, from the vantage point of geometric
structures
on
the jet space of solutions, one can see only a handful of ``special
features''
of the vacuum equations.  Still, let us summarize our results and what
they
tell us about the vacuum equations.

The generalized symmetries include a constant scale symmetry and a
diffeomorphism symmetry.  The scale symmetry simply indicates that there
are no length scales set by the vacuum equations; this symmetry is
absent
if
one modifies the equations using dimensionful constants, {\it e.g.}, if
one
includes a cosmological term in the equations.   The diffeomorphism
symmetry reflects the general covariance of the Einstein equations.  We
expect this symmetry to be present in any generally covariant system of
field equations.  Aside from these well-known transformations, the
Einstein
equations are devoid of symmetry.

The only closed--not--exact form locally constructed from Ricci-flat
metrics
corresponds to a topological conservation law---the conservation of
``kink
number''.   This conservation law reflects the non-trivial topology of
the
bundle of Lorentzian metrics over spacetime and will arise in any system
of field equations for a Lorentzian metric.  If the metric is
Riemannian,
this conservation law is absent.  The absence of any other conserved
3-forms can
be traced back to the absence of suitable generalized symmetries.  But
this
does not explain the dearth of lower-degree conservation laws.

Remarkably, it is possible to give a rather simple theory of
lower-degree conservation laws in a general field theory \cite{Anderson
and Torre (1995)}, which can be
thought of as somewhat analogous to Noether's theory of conserved
currents \cite{Olver (1993)}.  With some mild technical assumptions it
is possible to show
that in order for a set of Lagrangian field equations to admit
lower-degree
conservation laws two conditions must be met.  First, the theory must be
a
``gauge theory'', that is, it must admit some form of {\it gauge
transformation}, where we define a gauge transformation as a generalized
symmetry built from arbitrary functions of spacetime.  Second, the
solutions to the field equations must be such that they always allow for
{\it
gauge symmetries}, that is, there always exists a gauge transformation
that
leaves each solution invariant\footnote{The gauge symmetry of a solution
will in
general vary with the choice of solution.}.  Thus, the gauge
transformation
of the Einstein equations is the diffeomorphism symmetry ((17) with
$c=0$),
and a gauge symmetry of a solution $g_{ab}$ to the field equations would
be a diffeomorphism which does not change that solution.  The
infinitesimal
gauge
symmetry is then generated by a {\it generalized Killing vector field},
that is, a
vector field locally constructed from the
metric and its derivatives to some order which satisfies the Killing
equations when the metric is Ricci-flat.  The generic solution to
the
vacuum equations admits no Killing vector fields.  More precisely, it is
possible to show that there are no generalized Killing vector fields,
and
so
we can say that
the absence of lower-degree conservation laws for the Einstein equations
reflects the absence of isometries of generic solutions.  If we consider
reductions of the Einstein equations obtained by demanding the solutions
always admit a Killing vector, then the general theory
leads to lower-degree conservation laws such as shown in (20) and (21).

Finally, we have classified generalized conservation laws built locally
from solutions to the linearized equations and found only the symplectic
current and its ``potential''.  These conservation laws reflect the
variational
properties of the Einstein equations.  As is well-known \cite{Ashtekar
(1990)}, a conserved
symplectic current arises for any field equations derivable from a
Lagrangian. Thus the conserved 3-form (23) reflects the fact that the
Einstein
equations can be derived from the Einstein-Hilbert Lagrangian.
The essential uniqueness of the pre-symplectic current leads to a
uniqueness result for variational principles for the vacuum Einstein
equations, which will be presented elsewhere.
Normally,
the current defining the symplectic potential for a system of Lagrangian
field equations is not conserved.  However, it is not hard to show that
the
current defining the symplectic potential {\it is} conserved provided
the
Lagrangian can be chosen to vanish when the field equations hold.  Thus
the closed 3-form (26) reflects the fact that the Einstein-Hilbert
Lagrangian vanishes on Ricci-flat metrics.

In this article I have discussed structural features of the vacuum
Einstein
equations which can be uncovered using spinor--jet space techniques.
The
techniques that were used can be generalized to analyze
related systems of equations, specifically (i) reductions of the
Einstein
equations such as obtained by restricting attention to solutions with
one
or
more Killing vector fields; and (ii) the Einstein equations with matter
couplings, {\it e.g.}, the
Einstein-Maxwell equations.  As mentioned above, we expect non-trivial
lower-degree conservation laws to arise when one analyzes the
Einstein-Killing
equations, and a classification of such conservation laws is currently
in
progress.  Matter couplings can also induce lower degree conservation
laws; for example, the Einstein-Maxwell equations admit a 2-form
conservation law.  More intriguing, if perhaps on somewhat less firm
physical ground, are matter couplings dictated by Kaluza-Klein
reductions
of higher-dimensional vacuum relativity.  For example, the classical
reduction from five to four dimensions corresponds to an
Einstein-Maxwell-scalar field theory.  This reduction is dictated by the
assumption that the five-dimensional vacuum theory admits a Killing
vector
field, which, on general grounds, indicates the existence of
``characteristic
cohomology''.  The details of such investigations will be presented
elsewhere.

\head 5\enspace References/Bibliography\endhead

\enddocument